\input epsf
\documentstyle[preprint,eqsecnum,aps]{revtex}
\tightenlines

\def\INSERTCAP#1#2{\vbox{%
{\narrower\noindent%
\multiply\baselineskip by 3%
\divide\baselineskip by 4%
{\rm Table #1. }{\sl #2 \medskip}}
}}

\count255=\time\divide\count255 by 60 \xdef\hourmin{\number\count255}
  \multiply\count255 by-60\advance\count255 by\time
  \xdef\hourmin{\hourmin:\ifnum\count255<10 0\fi\the\count255}

\begin{document}

\draft
\preprint{\vbox{
\hbox{UCSD/PTH 97-28} \hbox{JLAB-THY-97-44}
}}

\title{Consistency constraints on $m_s$ from QCD dispersion relations
and chiral perturbation theory in $K_{\ell 3}$ decays}

\author{Richard F. Lebed\footnote{lebed@jlab.org}}

\address{Thomas Jefferson National Accelerator Facility\\ 12000
Jefferson Avenue, Newport News, VA 23606}

\author{Karl Schilcher\footnote{Permanent address: Institut f\"{u}r
Physik, Johannes Gutenberg-Universit\"{a}t, Staudingerweg 7,
D-55099 Mainz, Germany}}

\address{Department of Physics, University of California at San
Diego\\ La Jolla, CA 92093}

\date{October, 1997}

\maketitle
\begin{abstract}
	We use both old and new theoretical developments in QCD
dispersion relation constraints on the scalar form factor in the decay
$K \to \pi \ell \nu_{\ell}$ to obtain constraints on the strange quark
mass.  The perturbative QCD side of the calculation incorporates up to
four-loop corrections, while the hadronic side uses a recently
developed parameterization constructed explicitly to satisfy the
dispersive constraints.  Using chiral perturbation theory ($\chi$PT)
as a model for soon-to-be measured data, we find a series of lower
bounds on $m_s$ increasing with the accuracy to which one believes
$\chi$PT to represent the full QCD result.

\end{abstract}



\narrowtext

\section{Introduction}

	The level of interest in using dispersion relations to study
the analytic properties of Green functions has ebbed and flowed for
decades.  Recently, attention has increased due to studies of bounds
obtained on form factors from heavy hadron semileptonic decays, which
can be used in concert with Heavy Quark Effective Theory to isolate
the CKM elements $|V_{cb}|$ and $|V_{ub}|$ (see \cite{BGL97} for a
compilation of references).  Such studies were initially motivated by
the 1981 paper of Bourrely, Machet, and de Rafael\cite{BMR}, which
first applied the basic method of equating the dispersion integral
over hadronic form factors to its perturbative QCD evaluation in the
deep Euclidean region, in that paper for the case of $K_{\ell 3}$
decays.  Here we revisit $K_{\ell 3}$ decays, armed with new
improvements in technique and the promise of vastly more data to be
produced at DA$\Phi$NE\cite{DAFNE}.  Because the dispersive bound uses
QCD to constrain the shape of each hadronic form factor (as a function
of momentum transfer), one can use the experimentally measured form
factors to obtain limits on QCD parameters such as the quark masses;
however, until this data is available, one can use precisely the same
method to take the form factor from a given model or calculation as
being correct, and discover limits on values of the QCD parameters for
which the shape of the form factor is consistent with the dispersive
bounds.\footnote{Alternately, one could insist that particular QCD
inputs are correct and use the bounds to obtain limitations on the
consistency of the model with QCD.  Since we are using the model as a
proxy for real data, we do not follow this approach here.}

	In particular, we use the one-loop corrected chiral
perturbation theory ($\chi$PT) result of Gasser and Leutwyler\cite{GL}
for the scalar form factor in $K_{\ell 3}$ decays.  The scalar form
factor is especially interesting because it contributes to the scalar
current correlator, which is proportional to $(m_s - m_u)^2$ ({\it
i.e.}, vanishing for flavor-conserving processes), and is thus more
sensitive to quark mass values.  Moreover, it satisfies low-energy
theorems in $\chi$PT to a high degree of precision.  Our results may
also be used for the vector form factor to obtain interesting
information, but our thrust in this paper is to show how experiment
will be able to teach us about limits on $m_s$, and specifically, that
the consistency of $\chi$PT and QCD places substantial lower limits on
the strange quark mass.

	This paper is organized as follows: In Sec.~II we summarize
the most important parts of the dispersion relation approach with
references to papers explaining the fine points in more detail, as
well as describe our $\chi$PT inputs.  Section~III presents the QCD
result for the scalar current correlator to four loops, including
leading nonperturbative effects.  In Sec.~IV we present improved
versions of the analytic functions used to parameterize form factors
consistent with QCD and describe how to judge the quality of fits to
this form.  In Sec.~V we present our results and conclude.

\section{Using the Dispersion Relations and {\boldmath $\chi$}PT} 

	 By means of Cauchy's theorem, the dispersion integral simply
relates an integral of a Green function (in our case, a current
two-point correlator) expressed over a kinematic region of momentum
transfer $q^2$ where hadronic quantities conveniently describe the
physics, to the Green function itself evaluated at a point deep in the
Euclidean region, where perturbative QCD is reliable.  In the present
case, the current is $V^\mu = \bar s \gamma^\mu u$, whose divergence
$\partial \cdot V = i ( m_s - m_u ) (\bar s u)$ is just the scalar
current multiplied by an explicit factor of quark masses; it is this
feature that leads us to interesting bounds on $m_s$.  The scalar
correlator, related to the vector correlator by means of a Ward
identity, is given by
\begin{equation} \label{corr}
\psi(Q^2) = i \int d^4 x \, e^{iqx} \langle 0 | {\rm T}
\partial_\mu V^\mu (x) \partial_\nu V^\nu (0)^\dagger | 0 \rangle ,
\end{equation}
where $Q^2 \equiv -q^2$.  The unsubtracted dispersion relation reads
\begin{equation} \label{disp}
\psi(Q^2) = \frac{1}{\pi} \int_0^\infty dt \, \frac{{\rm Im} \,
\psi(t)}{(t+Q^2)} .
\end{equation}
The $n$-times subtracted form of (\ref{disp}) is readily obtained by
taking $Q^2$ derivatives on both sides, and in the remainder of this
section we take $\psi(Q^2)$ to mean the perturbatively computed
two-point function with any fixed number of subtractions.

	The l.h.s.\ of (\ref{disp}) is evaluated at a point $Q^2$ far
from the resonant region of $V^\mu$, which for $s \to u$ transitions
is satisfied when $Q^2 \gg \Lambda^2_{\rm QCD}$.  In practice, we
choose $Q^2 = 4$ GeV$^2$, which is popular in lattice simulations.
The r.h.s.\ of (\ref{disp}) is expressed as a sum of integrals over
hadronic quantities.  It is a manifestly nonnegative quantity, and so
neglecting any subset of contributions produces a strict inequality,
in that any subset of hadronic form factors in any kinematic region
connected by crossing symmetry is limited by an expression computable
in QCD.  In particular, integrals of the form factors for $K \to \pi$
in the crossed kinematic region of (vacuum $\to \bar K \pi$) are
bounded, leading to restrictions on the shape of form factors
permitted by QCD.  The tightness of these bounds is of course
regulated by numerical inputs to the QCD calculation, and particularly
in this case by $m_s$.  It is precisely this sensitivity that is
useful to us: Given the shape of a $K \to \pi$ form factor obtained
from data or a model, one may ask which values of $m_s$ permit this
shape to be consistent with QCD.

	The derivation of the bounds on the shapes of semileptonic
decay form factors and its expression in terms of a well-defined
parameterization draws upon techniques developed in Refs.~\cite{BMR},
and \cite{BL,BGLAug,BGLApr}, and the approach is explained in detail
in \cite{BGLAug}.  Here we present expressions only to establish
notation for later use.  Beginning with the kinematic points
\begin{equation}
t_{\pm} \equiv (m_K \pm m_\pi )^2 ,
\end{equation}
it is convenient in such decays to use the kinematic variable
\begin{equation} \label{zdef}
z(t;t_s) \equiv \frac{\sqrt{t_+ - t} - \sqrt{t_+ - t_s}}{\sqrt{t_+ -
t} + \sqrt{t_+ - t_s}} ,
\end{equation}
where $t_s < t_+$ is a kinematic scale fixed for convenience, as
described below.  For each form factor $F(t)$ there is a computable
function $\phi_F (t;t_s;Q^2)$ obtained from (\ref{disp}), and in terms
of these quantities all functional forms for the form factors allowed
by QCD may be parameterized by
\begin{equation} \label{masterparam}
F(t) = {\sqrt{\psi(Q^2)} \over \phi_F (t;t_s;Q^2) } \sum_{n=0}^\infty
                   a_n \, z(t;t_s)^n \ \ ,
\end{equation}
where the coefficients $a_n$ are unknown parameters obeying
\begin{equation} \label{abound}
 \sum_{n=0}^\infty  |a_n|^2 \le 1 \ .
\end{equation}
This result employs analyticity of the two-point function away from
hadronic thresholds, crossing symmetry between the matrix elements for
$\bar K \pi$ production and $K \to \pi$, and knowledge of QCD at
$Q^2$.  To lowest order in $G_F$ this form is exact, for there are no
resonant poles, multiparticle continuum states, or anomalous
thresholds in the scalar channel below the threshold $t = t_+$ of
(vacuum $\to \bar K \pi$).  To be precise, one must also distinguish
the two isospin channels (vacuum $\to \bar K^0 \pi^-$) and (vacuum
$\to K^- \pi^0$).  Apart from a small explicit isospin breaking in the
form factors, whose inclusion is described below, the two channels
have slightly different values of $t_{\pm}$.  Since nonanalytic $q^2$
behavior in the correlator, which arises from physical and anomalous
particle thresholds, is what determines the form of the
parameterization (\ref{masterparam}), we take $t_+$ to be the smaller
of the two possibilities, $(m_{K^+} + m_{\pi^0})^2$, so that no
threshold is ever crossed in the region where (\ref{masterparam}) is
used.

	The two form factors relevant to $K_{\ell 3}$ decays are
defined by
\begin{equation} \label{ff}
\langle \pi (p') | V^\mu | K (p) \rangle = 2f_+(q^2) \left( p^\mu -
\frac{p \cdot q}{q^2} q^\mu \right) + d(q^2) \frac{q^\mu}{q^2}
\end{equation}
where $q^\mu = (p-p')^\mu$.  This decomposition separates the vector
$f_+$ and scalar $d$ form factors, and consequently the bounds studied
in this work apply to $d$.  In the differential width, $|d(q^2)|^2$
appears multiplied by a helicity suppression factor of $m_\ell^2$.

	In lieu of very precise experimental information, we use the
results\cite{GL} of one-loop chiral perturbation theory for the scalar
form factor as ``data''.  At two kinematic points in particular, $q^2
= 0$ and $q^2 = (m_K^2 - m_\pi^2)$, predictions of amazing precision
have been made.  At $q^2 = 0$, the authors of \cite{GL} find
\begin{equation} \label{AGf}
f_+^{K^0 \pi^-} (0) = 0.977 , \hspace{2em} f_+^{K^+ \pi^0} (0) = 0.998
.
\end{equation}
By the Ademollo-Gatto theorem\cite{AG}, the symmetry relation $f_+ (0)
= 1$ is violated by terms of order $m_s^2$, and we see these
percent-level deviations in the predictions (\ref{AGf}).
Uncertainties on (\ref{AGf}) are even smaller, perhaps a few parts per
$10^3$.  The relation to the scalar form factor at this point is
\begin{equation}
d(0) = (m_K^2 - m_\pi^2) f_+ (0) .
\end{equation}
At $q^2= (m_K^2 - m_\pi^2)$, the value of $d(q^2)$ is given by the
Callan-Treiman relation\cite{CT},
\begin{equation}
d(m_K^2 - m_\pi^2) = (m_K^2 - m_\pi^2) \left( \frac{f_K}{f_\pi} +
\Delta_{\rm CT} \right) ,
\end{equation}
where $f_K /f_\pi = 1.22 \pm 0.01$, and $\Delta_{\rm CT}$ measures
deviations from the exact Callan-Treiman limit.  As calculated in
\cite{GL}, $\Delta_{\rm CT} = -3.5 \cdot 10^{-3}$.  Since no strange
$0^+$ resonances appear until the relatively high mass $K^*_0$(1430)
and $K^*_0$(1950), the form factor in the decay region should be very
smooth (and indeed, almost linear) to a very good approximation.  It
is therefore very reasonable to assume that the scalar form factor is
given in the intermediate region, {\it i.e.,} between $q^2 = 0$ and
$q^2 = (m_K^2 - m_\pi^2)$, by chiral perturbation theory to the same
accuracy as at the endpoints.  It is, of course, difficult to estimate
this error, although some indication may be obtained from a
forthcoming two-loop $\chi$PT calculation\cite{PS}, despite the
possible appearance there of currently undetermined renormalization
constants.  At any rate, since the one-loop corrections to the
low-energy theorems are already quite small, the whole framework of
$\chi$PT would crumble if new experiments showed deviations from the
predictions by more than a few percent.

\section{The Scalar Correlator in QCD}

	Instead of the divergent scalar correlator $\psi (Q^2)$, we
consider its second derivative $\psi^{\prime\prime} (Q^2)$, which is
free of renormalization point dependence and therefore satisfies a
homogeneous renormalization group (RG) equation.  $\psi^{\prime\prime}
(Q^2)$ is calculated in perturbative QCD for large $Q^2$ using an
expansion in powers of the quark masses and the operator product
expansion.  The leading term has been calculated to four
loops\cite{Chet1}, and the $O(m_s^2/Q^2)$ correction to three
loops\cite{Chet2}.  The $O(m_s^4/Q^4)$ contribution is best considered
together with the nonperturbative condensate terms.  Collecting these
results,
\begin{eqnarray} \label{QCD}
\psi^{\prime\prime} (Q^2) & = & \frac{6(m_s-m_u)^2}{(4\pi)^2 Q^2}
\left\{ 1 + \frac{11}{3} \left( \frac{\alpha_s}{\pi} \right)  +
\left( \frac{\alpha_s}{\pi}
\right)^2 \left( \frac{5071}{144} - \frac{35}{2} \zeta(3) \right)
\right. \nonumber \\ & & \left. \hspace{1em}
+ \left( \frac{\alpha_s}{\pi} \right)^3 \left[ - \frac{4781}{9} +
\frac{1}{6} \left( \frac{4748953}{864} - \frac{\pi^4}{6} -
\frac{91519}{36} \zeta(3) + \frac{715}{2} \zeta(5) \right) +
\frac{475}{4} \zeta(3) \right] \right\} \nonumber
\\ & &
-\frac{12 (m_s - m_u)^2 m_s^2}{(4\pi)^2 Q^4} \left\{ 1 + \frac{28}{3}
\left( \frac{\alpha_s}{\pi} \right) + \left( \frac{\alpha_s}{\pi}
\right)^2 \left( \frac{8557}{72} - \frac{77}{3} \zeta(3) \right)
\right\} \nonumber \\ & &
+ \frac{(m_s - m_u)^2}{Q^6} \left\{ 2 \langle m_s \bar u u \rangle
\left[ 1 + \frac{23}{3} \left( \frac{\alpha_s}{\pi} \right) \right] -
\frac{1}{9} I_G \left[ 1 + \frac{121}{18} \left( \frac{\alpha_s}{\pi}
\right) \right] \right. \nonumber \\ & & \left. \hspace{6em}
+ I_s \left[ 1 + \frac{64}{9} \left( \frac{\alpha_s}{\pi} \right)
\right] - \frac{3}{7\pi^2} m_s^4 \left[ \left( \frac{\pi}{\alpha_s}
\right) + \frac{155}{24} \right] \right\} ,
\end{eqnarray}
where the RG-invariant condensate combinations $I_s$ and $I_G$ are
given by
\begin{eqnarray}
I_s & = & m_s \langle \bar s s \rangle + \frac{3}{7\pi^2} m_s^4 \left[
\left( \frac{\pi}{\alpha_s} \right) - \frac{53}{24} \right] , \\
I_G & = & -\frac{9}{4} \left< \frac{\alpha_s}{\pi} G^2 \right> \left[
1 + \frac{16}{9} \left( \frac{\alpha_s}{\pi} \right) \right] + 4 \left(
\frac{\alpha_s}{\pi} \right) \left[ 1 + \frac{91}{24} \left(
\frac{\alpha_s}{\pi} \right) \right] m_s \langle \bar s s \rangle
\nonumber \\ & & +
\frac{3}{4\pi^2} \left[ 1 + \frac{4}{3} \left( \frac{\alpha_s}{\pi}
\right) m_s^4 \right] ,
\end{eqnarray}
where we have set $n_f = 3$ and omitted logarithms that vanish when
taking $\mu = Q \equiv \sqrt{Q^2}$.  Note that the peculiar
$\pi/\alpha_s$ terms cancel.  The full result depends logarithmically
on the renormalization point $\mu$ and on the parameters of the
theory, like $\alpha_s$, $m_s$, and condensates, which are
renormalized at $\mu$; however, as has been advocated in~\cite{CPS},
we implement the RG improvement for the case of the scalar correlator
in the following way: $\psi^{\prime\prime} (Q^2)$ is evaluated at $\mu
= Q$, and the parameters $\alpha_s$ and $m_s$ are extrapolated from a
chosen reference point (in our case $\Lambda_{\overline{\rm MS}})$ to
$\mu = Q$ using the four-loop beta functions (compiled in \cite{CPS}).
The condensates are so poorly known and their effect at the chosen
scale $Q^2 = 4$ GeV$^2$ so small that their $\mu$ dependence may be
ignored.

	The numerical values chosen for the QCD inputs are $Q^2 = 4$
GeV$^2$, $\Lambda^{n_f = 3}_{\overline{\rm MS}} = 380 \pm 60$, $m_u =
m_s/25$, $\left< m_s \bar u u \right> = \left< m_s \bar s s \right> =
-f_K^2 m_K^2 = -0.031$ GeV$^4$, and $\left< \alpha_s G^2 / \pi \right>
= $ 0.02--0.06 GeV$^4$, although the sensitivity of the analysis to
these nonperturbative parameters is small.  The scale $Q$ is arbitrary
subject to the constraints that if it is too small, perturbative QCD
is unreliable, while if it is too large, the dispersive bounds thus
obtained are weak.

\section{Parameterization and Quality of Fit}

	In order to use the parameterization of
Eq.~(\ref{masterparam}), one requires expressions for the function
$\phi$ for each form factor.  In notation designed to be similar to
that of \cite{BMR}, we define
\begin{equation}
\beta_0 \equiv \sqrt{\frac{t_+}{t_+ - t_s}} , \hspace{2em} 
\beta_1 \equiv \sqrt{\frac{t_+ - t_-}{t_+ - t_s}} , \hspace{2em}
\beta_2 \equiv \sqrt{\frac{t_+ + Q^2}{t_+ - t_s}} .
\end{equation}
Note that, in our notation, one must set the parameter $t_s = -Q^2$ to
obtain the limit of Ref.~\cite{BMR}, in which case $\beta_2 = 1$.  The
$\phi$ for each form factor defined in Eq.~(\ref{ff}) is given by
\begin{eqnarray}
\phi_d (z) & = & \sqrt{\frac{\eta^2}{2\pi}} \frac{(1+z)}{(1-z)^2}
\frac{1}{t_+ - t_s} \left[ \beta_0 + \frac{1+z}{1-z} \right]^{-1}
\left[ \beta_1 + \frac{1+z}{1-z} \right]^{1/2} \left[ \beta_2 +
\frac{1+z}{1-z} \right]^{-3} , \\
\phi_{f_+} (z) & = & \sqrt{\frac{\eta^2}{48\pi}} \frac{(1+z)^2}
{(1-z)^3}
\sqrt{\frac{Q^2}{t_+ - t_s}} \left[ \beta_0 + \frac{1+z}{1-z}
\right]^{-3} \left[ \beta_1 + \frac{1+z}{1-z} \right]^{3/2}
\left[ \beta_2 + \frac{1+z}{1-z} \right]^{-2} .
\end{eqnarray}
These expressions for $\phi$ differ from those in \cite{BGL97},
because the dispersion integrals are formulated using different linear
combinations of the two polarization tensor component functions than
used in the other work, leading to a different pattern of
subtractions.  The factors $\eta$ represent Clebsch-Gordan
coefficients based on isospin symmetry; that is, both charged and
neutral $K$ processes contribute to the dispersion relation, so we
exploit the near equality of their contributions.  For the decay $K_L
\to \pi^{\pm} \ell^{\mp} \nu_\ell$ (whose width equals, by CPT, either
that for $K^0$ or $\bar K^0$ semileptonic decays), $\eta^2 = 3/2$,
while for $K^+ \to \pi^0 \ell^+ \nu_\ell$, $\eta^2 = 3$.  If isospin
breaking between charged and neutral $K$ processes is significant, one
can incorporate this difference into $\eta^2$, or even include a
factor with $q^2$ dependence (written in terms of $z$) if the pattern
of isospin breaking is known.  In practice, we obtain a conservative
correction to $\eta_{K^+}^2 = 3$ by supposing that the form factor
ratio represented by (\ref{AGf}) at $q^2 = 0$ persists for all values
of $q^2$ in the $\bar K \pi$ production region, to obtain an effective
$\eta_{K^+}^2 = 2.92$.

	In the original work \cite{BMR}, the bounds are expressed as
the determinant of an $n \times n$ semipositive-definite matrix, where
the form factor is assumed known at $(n-2)$ points.  This produces an
envelope of allowed form factors resembling a chain of sausages, since
the form factor is required to pass through each of the $(n-2)$
points.  Although presented explicitly in \cite{BMR} for only 1 or 2
fixed points, the determinant method can of course be generalized to
an arbitrary number of points, with a corresponding increase in the
complexity of algebraic expressions appearing (see, {\it e.g.},
\cite{Lel}).  However, once the envelope of points allowed by the
dispersion relation and chosen fixed points is established, it is {\em
not\/} true that any curve lying within this envelope still satisfies
the dispersive bound.  In contrast, the parameterization of
Eq.~(\ref{masterparam}) subject to (\ref{abound}) always satisfies the
determinant bound of arbitrarily high degree.

	The central features that make the parameterization
(\ref{masterparam}) useful are the bound (\ref{abound}) on the
coefficients $a_n$ and the smallness of the kinematic variable $z$
over the full range for allowed semileptonic decay.  It follows that
one may express the form factor over the full range using only the
first $N+1$ parameters $\{ a_0, a_1, \ldots , a_N \}$, with the
remaining infinite set bounded in magnitude and forming a theoretical
{\em truncation error}\cite{BGLApr}, $\delta_N$:
\begin{equation} \label{trunc}
\delta_N \equiv \frac{\sqrt{\psi^{\prime\prime} (Q^2)}}{|\phi(z)|}
\sqrt{1 - \sum_{n=0}^N |a_n|^2} \, \frac{|z|^{N+1}}
{\sqrt{1-|z|^2}} ,
\end{equation}
which means that the form factor fit to these parameters using
(\ref{masterparam}) has a {\em theoretical\/} uncertainty no larger
than $\delta_N$.

	The kinematic parameter $t_s$ is used to minimize the size of
this already small truncation error\cite{BGL97,BL}.
Equation~(\ref{trunc}) makes it clear that this minimization occurs
when $z=0$ lies within the kinematic range chosen for the fit, $t \in
[t_{\rm min} , t_{\rm max}]$; from (\ref{zdef}) one sees that this can
occur only if $t_s$ also assumes some value in this range.  Therefore,
the truncation error is minimized by plotting (\ref{trunc}) as a
function of $t,t_s \in [t_{\rm min}, t_{\rm max}]$, and finding that
value of $t_s$ for which the maximum value over all $t \in [t_{\rm
min} , t_{\rm max}]$ is smallest.

	Now suppose that the ``data'', in our case the $\chi$PT
expression for the form factor, has an ``experimental'' uncertainty
$\Delta$.  In order to state that $\chi$PT ``data'' agrees with QCD to
within $\Delta$, it must be possible to expand the QCD fit to an order
$N$ such that $\sum_{n=0}^N |a_n|^2 \leq 1$ {\em and\/} $\delta_N \leq
\Delta$.  In other words, in order to be certain that the {\em exact
all-orders\/} QCD form factor $F_{\rm exact}$ (which is not accessible
to us) lies within an uncertainty $\Delta$ of the data $F_{\chi{\rm
PT}}$, we require that the deviation $\delta_N$ of the {\em
truncated\/} QCD form factor $F_{\rm trunc}$ from $F_{\rm exact}$ is
smaller than\footnote{If one prefers to combine the theoretical and
experimental uncertainties, then the requirement becomes $|F_{\rm
trunc} - F_{\rm \chi{\rm PT}}| < \sqrt{\Delta^2 - \delta_N^2}$.} the
experimental uncertainty $\Delta$.  This is the central argument of
our reasoning.

	The requirement that both $\delta_N \leq \Delta$ and
(\ref{abound}) are satisfied is the key to obtaining bounds on
parameters in $\psi^{\prime\prime} (Q^2)$, in particular $m_s$.  A
curve such as $F_{\chi{\rm PT}}$, which does not {\it \`{a} priori\/}
satisfy (\ref{abound}), when expanded as a power series in $z$
eventually produces a value of $a_N$ for some $N$ so large that
(\ref{abound}) is violated.  This does not necessarily mean that
$F_{\chi{\rm PT}}$ violates QCD, because it is only required to equal
the $F_{\rm exact}$ within an uncertainty of $\Delta$.  However, the
value of $N$ must be large enough that $\delta_N \leq \Delta$, or else
$F_{\rm trunc}$ expanded to $n=N$ is not necessarily a good enough
approximation to use in place of $F_{\rm exact}$; if it is not
possible to carry out this fit and maintain Eq.~(\ref{abound}), then
the chosen inputs are not consistent with QCD bounds.  When the fit is
successful, one concludes that the true QCD form factor is actually
given by the fit of the $\chi$PT form factor to the parameterization
up to order $N$, and the higher-order terms cannot be probed with
$\chi$PT since they lie within our stated uncertainty $\Delta$.

	The particular nature of the fit is irrelevant to us, since
one requires additional physical input to distinguish two curves that
otherwise satisfy all the requisite conditions.  For example, we use
(\ref{masterparam}) and choose to Taylor expand the $\chi$PT form
factor in $z$ as an expression for $F(z) \phi_F(z) /
\sqrt{\psi^{\prime\prime} (Q^2)}$, to obtain $a_n$ values until
(\ref{abound}) is violated and throw away all higher $a_n$.
Alternately, one may choose a highest-order $N$ and perform a $\chi^2$
fit to $\{ a_0, a_1, \ldots , a_N \}$ subject to the constraint
(\ref{abound}); this would be the preferred method of analysis with
binned data.

	Now consider the dependence of the bounds on the strange quark
mass, which enters approximately linearly in
$\sqrt{\psi^{\prime\prime} (Q^2)}$ as seen in (\ref{QCD}).  If one
fixes all other quantities and decreases $m_s$, it is clear from
(\ref{masterparam}) that one obtains the same form factor by
increasing each $a_n$ by the same factor.  However, eventually the
$a_n$'s are large enough to saturate (\ref{abound}), and thus one
finds a minimal allowed value for $m_s$.  There is no corresponding
maximal value, for one can certainly make each $a_n$ as small as
desired and still satisfy (\ref{abound}).  Furthermore, as we increase
the precision $\Delta$ to which we believe the $\chi$PT form factor
holds, $\delta_N$ must decrease, so that we must go to a higher order
$N$ in the parameterization expansion.  It becomes increasingly more
difficult to avoid large values of $a_n$ unless $m_s$ is increased, so
the lower bound on $m_s$ becomes larger as $\Delta$ decreases.

\section{Results and Conclusions}

	We present lower bounds on $m_s \equiv m_s^{\overline{\rm
MS}}$(1 GeV) from the form factor $F(t) \equiv d(t)/(m_K^2-m_\pi^2)$
of $K^+ \to \pi^0 \ell^+ \nu_\ell$, since its bound (due to the
isospin factor $\eta$) is roughly $\sqrt{2}$ tighter than that from
$K_L$ decay, as described above.  The natural tendency for the
truncation errors $\delta_N$ is to decrease an order of magnitude with
each unit of $N$, since $z_{\rm max} \approx -z_{\rm min} \approx 0.1$
for optimal values of $t_s$ over the range $t_{\rm min} = 0$, $t_{\rm
max} = (m_K^2 - m_\pi^2)$; however, this is somewhat counteracted by
the fact that $\delta_N \propto \sqrt{\psi^{\prime\prime}}$, which
increases when the smallest allowed value of $m_s$ increases with $N$.
Moreover, $\delta_N$ can be much smaller than the numbers we present,
since we do not include the effects of the $\sqrt{1 - \sum_n |a_n|^2}$
term in (\ref{trunc}); near the saturation of (\ref{abound}) by the
first few terms, this is a large suppression.

	Our numerical results are summarized in Table~I.  The
conclusion we draw is that, if the $\chi$PT form factor calculation is
believed only to a precision of 2--5\%, then we conclude only that
$m_s > 40$ MeV.  At a precision of $\frac 1 2$--1\%, $m_s > 90$ MeV,
and at 1/20\%, $m_s > 140$ MeV.  The same sort of analysis will be
possible with data, although the details of the fit will differ
somewhat.

	We conclude by comparing briefly with other recent
approaches\cite{LRT,Ynd,Jam} that bound $m_s$ using dispersion
relations.  Typically, what is done is to saturate the hadronic side
as much as possible with phenomenological input from expected pole
contributions, and/or a continuum contribution in the deep Minkowski
region of the hadronic integral modeled by the perturbative QCD
result.  Often, multiple constraints are obtained by taking moments of
both sides.  We view this work as complementary to those approaches.
On one hand, it is minimal in the sense that only $\bar K \pi$, which
is presumably but a small portion of the total hadronic result, is
used in the bounding inequalities; of course, modelers may obtain
stronger bounds by including $\bar K^*_0$ poles or continuum
contributions at the cost of introducing model-dependent inputs.
Although we considered only one moment in our calculation, certainly
additional moments also give constraints, although then $Q^2$ must be
adjusted to larger values in order to make sure that $\psi^{(n)}
(Q^2)$ is calculable perturbatively.  On the other hand, it includes
data from semileptonic form factors measured over their entire
kinematic range, and so is expected to provide a substantial amount of
additional input to such constraints.  We expect that combined program
of calculations from all of these approaches will deliver rather
strong constraints on $m_s$.

\bigskip
\vbox{\medskip
\hfil\vbox{\offinterlineskip
\hrule
\halign{&\vrule#&\strut $\, \,$ \hfil$#$ $\, \,$ \hfil\cr
height0pt&\omit&&\omit&&\omit&&\omit&&\omit&&\omit&&\omit&\cr
& N  && \{ a_n \} && t_s ({\rm GeV}^2)  && \psi^{\prime\prime} \cdot
10^5 && \delta_{N, \, {\rm max}} && \Delta F_{\rm max} &&
m_s ({\rm MeV}) & \cr
\noalign{\hrule}
& 1 && a_0 = +0.910  && 0.144 && 1.16  && 1.4 \cdot 10^{-2}
&& 2.8 \cdot 10^{-2} &&  > 41 &\cr 
&   && a_1 = +0.414  &&       &&       &&
&&                   &&       &\cr 
\noalign{\hrule}
& 2 && a_0 = +0.418  && 0.142 && 5.54  && 3.1 \cdot 10^{-3}
&& 3.4 \cdot 10^{-3} && > 90  &\cr
&   && a_1 = +0.184  &&       &&       &&
&&                   &&       &\cr
&   && a_2 = -0.890  &&       &&       &&
&&                   &&       &\cr
\noalign{\hrule}
& 3 && a_0 = +0.273  && 0.140 && 13.01 && 5.0 \cdot 10^{-4}
&& 1.5 \cdot 10^{-4} && > 139 &\cr
&   && a_1 = +0.118  &&       &&       &&
&&                   &&       &\cr
&   && a_2 = -0.584  &&       &&       &&
&&                   &&       &\cr
&   && a_3 = -0.756  &&       &&       &&
&&                   &&       &\cr } \hrule} \hfil
\medskip
\\
\INSERTCAP{1}{Bounds on $m_s \equiv m_s^{\overline{\rm MS}}$ (1 GeV),
fit parameters, and truncation errors based on the saturation
(\ref{abound}) of the parameterization (\ref{masterparam}). $\Delta F$
is an abbreviation for $|F_{\rm trunc} - F_{\rm \chi{\rm PT}}|$.  The
numerical value of $\delta_{N, \, {\rm max}}$ here neglects the
$\sqrt{1-\sum_n |a_n|^2}$ term in (\ref{trunc}), which can be a large
suppression when (as here) (\ref{abound}) is nearly saturated by the
first $N$ terms.}}

\vskip 0.5in
{\it Acknowledgments}
\hfil\break
We thank the Theory Group of the University of California San Diego
for their hospitality, where much of this work was done, and
especially Ben Grinstein for comments on the manuscript.  RFL
acknowledges the support of Department of Energy under contracts Nos.\
DOE-FG03-97ER40506 and DE-AC05-84ER40150.  KS acknowledges partial
support by the Volkswagen-Stiftung.


\begin{thebibliography}{199}

\bibitem{BGL97}C. G. Boyd, B. Grinstein, and R. F. Lebed,
hep-ph/9705252 (to appear in Phys.\ Rev.\ D).

\bibitem{BMR}C. Bourrely, B. Machet, and E. de Rafael, Nucl.\ Phys.\ B
{\bf 189} (1981) 157.

\bibitem{DAFNE}J. Bijnens, G. Colangelo, G. Ecker, and J. Gasser, in
{\it The Second DA$\Phi$NE Physics Handbook}, Vol.\ I, Ed.\ by
L. Maiani, G. Pancheri, and N. Paver, INFN, Frascati, Italy (1995),
p. 315.  One year of running at DA$\Phi$NE is predicted to produce
3000 times more $K^+ \to \pi^0 \mu^+ \nu_\mu$ and 70 times more $K_L
\to \pi^{\pm} \mu^{\mp} \nu_\mu$ than in the current world sample.

\bibitem{GL}J. Gasser and H. Leutwyler, Nucl.\ Phys.\ B {\bf 250}
(1985) 517.

\bibitem{BL}C. G. Boyd and R. F. Lebed, Nucl.\ Phys.\ B {\bf 485}
(1997) 275.


\bibitem{BGLAug}C. G. Boyd, B. Grinstein, and R. F. Lebed, Nucl.\
Phys.\ B {\bf 461} (1996) 493.

\bibitem{BGLApr}C. G. Boyd, B. Grinstein, and R. F. Lebed, Phys.\
Lett.\ B {\bf 353} (1995) 306.


\bibitem{AG}M. Ademollo and R. Gatto, Phys.\ Rev.\ Lett.\ {\bf 13}
(1964) 264.

\bibitem{CT}C. G. Callan and S. B. Treiman, Phys.\ Rev.\ Lett.\ {\bf
16} (1966) 153.

\bibitem{PS}P. Post and K. Schilcher, in preparation.

\bibitem{Chet1}K. G. Chetyrkin, Phys.\ Lett.\ B {\bf 390} (1997) 309.

\bibitem{Chet2}K. G. Chetyrkin, in preparation.  For the full
results including renormalization point dependence, refer to
\cite{CPS}.

\bibitem{CPS}K. G. Chetyrkin, D. Pirjol, and K. Schilcher, Phys.\
Lett.\ B {\bf 404} (1997) 337; see also \\
%
K. G. Chetyrkin, C. A.  Dominguez, D. Pirjol, and K. Schilcher,
Phys.\ Rev.\ D {\bf 51} (1995) 5090; \\
%
M. Jamin and M. M\"{u}nz, Z. Phys.\ C {\bf 66} (1995) 633.

\bibitem{Lel}L. Lellouch, Nucl.\ Phys.\ B {\bf 479} (1996), 353.

\bibitem{LRT}L. Lellouch, E. de Rafael, and J. Taron, CPT Marseille
Report No.\ CPT-97/P.3519 and University of Barcelona Report No.\
UB-ECM-PF 97/15, hep-ph/9707523 (unpublished).

\bibitem{Ynd}F. J. Yndur\'{a}in, NIKHEF Report No.\ NIKHEF-97-035,
hep-ph/9708300 (unpublished).

\bibitem{Jam}M. Jamin, University of Heidelberg Report No.\
HD-THEP-97-51, hep-ph/9709484 (unpublished).

\end{thebibliography}
\end{document}